# The Instantaneous Wall Viscosity in Pipe Flow of Power Law Fluids:

# Case Study for a Theory of Turbulence in Time-Independent Non-Newtonian Fluids

Trinh, Khanh Tuoc

Institute of Food Nutrition and Human Health

Massey University, New Zealand

*K.T.Trinh@massey.ac.nz*

## ABSTRACT

This paper presents a new theory of turbulence in time-independent non-Newtonian fluids. The wall layer is modelled in terms of unsteady exchange of viscous momentum between the wall and the main stream, following the classic visualisation of inrush-sweep-ejection/burst. The thickness of the wall layer is found to be the same for Newtonian and purely viscous non-Newtonian fluids, when normalised with the instantaneous wall parameters at the onset of bursting. The results indicate that the mechanisms of turbulence in Newtonian and time-independent fluids are identical when structural similarity relations in turbulence are based on phase-locked parameters linked with the development of secondary flows rather than on time-averaged wall parameters. This similarity analysis collapses the local critical instantaneous friction factor data of both Newtonian and non-Newtonian fluids at the point of bursting into a single curve. The method greatly simplifies the analysis of turbulent transport phenomena in non-Newtonian fluids.

Keywords: Turbulence, time-independent non-Newtonian, Power law, pipe flow, wall layer

**Introduction**

Non-Newtonian turbulent flow has applications in the processing of food, mineral, oil and polymer products. Early research efforts were directed towards defining an equivalent viscosity, which could be substituted into Newtonian correlations to give estimates of the friction factor. For example Weltmann (1956) used the limiting viscosity at "limiting shear rate" $\mu_\infty$. Alves et al . (1952) used a "turbulent viscosity" associated with measurements of equivalent viscosity in turbulent flow. Metzner and Reed (1955) used an "effective viscosity" to define a so called Meztner-Reed generalised Reynolds number $Re_g$ required to collapse all data of the friction factor f against the Reynolds number into the Newtonian curve. They argued that plots of f against $Re_g$ would also collapse the data for turbulent flow onto the Newtonian curves modelled by Nikuradse (1932). This was disproved by the subsequent data of Dodge (1959). Bogue (1961) and later on Edwards and Smith (1980) argued that the use of the apparent viscosity at the wall shear rate $\mu_w$ collapsed the non-Newtonian data onto the Newtonian curve. But the procedure only really worked for fluids in the measurements of Bogue (1961) and Eissenberg and Bogue (1964) with a relatively high value of behaviour index n>0.89 where departure from Newtonian behaviour was small anyway. The method failed to correlate friction factor for Carbopol solutions obtained by Dodge (1959)and Shaver (1957)at high concentrations and therefore lower values of n. Edwards and Smith blamed viscoelastic effects but Dodge had argued that Carbopol used to generate the data used by Edwards and Smith showed very little viscoelasticity.

Major efforts were made in the 1960s to analyse the problem. Bowen (1961) proposed a method of scale up for turbulent pipe flow that is useful if some data of turbulent flow is available for one pipe diameter. Tomita (1959) used the Nikuradse formula (Nikuradse, 1932) but redefined both the friction factor and the Reynolds number for power law fluids from similarity considerations (Table 1). Clapp (1961) followed the technique of Prandtl(1935) and Karman (1934) to derive the friction factor from a model of the velocity profile consisting of a turbulent core following the well-known log law (Millikan, 1939), Prandtl (1935)) and a laminar sub-layer (Prandtl, 1935) which he assumed has the same thickness in Newtonian and power law fluids. Experimental measurements of the velocity profile in turbulent pipe flow have never been accurate and repeatable enough to verify this key assumption. More importantly Metzner (Clapp, 1961) pointed out that the rheological data used by Clapp cover a lower range of shear rates than those encountered in the actual flow experiments. Since it is always dangerous to extrapolate non-Newtonian rheological data, this uncertainty in the values of the consistency coefficient K and the behaviour index n may explain why there is a small difference between the data presented by Clapp(1961) and others like Dodge (1959), Bogue (1961) and Yoo (1974)

Dodge & Metzner (1959) used dimensional analysis to extend Millikan's logarithmic law to time independent non-Newtonian fluids. The term time-independent means that the fluids do not thin (thixotropy) or thicken (rheopexy) with duration of shear and do not exhibit viscoelasticity. While Dodge and Metzner showed that a certain amount of drag reduction exists for time independent non-Newtonian fluids compared with Newtonian fluids at the same Reynolds drag reduction in the presence of viscoelasticity is much more pronounced as identified by Toms (1949) a distinction not made by their contemporary Shaver and Merrill (1959). A second distinctive feature of the Dodge and Metzner correlations is that they are expressed in terms of the behaviour index $n'$, the slope of the log-log plot between the shear stress τ and the flow function $(\Gamma = 8V/D)$ where V is the average flow velocity and D the pipe diameter and not the index n which is the slope of the log-log plot between τ and the shear rate $\dot{\gamma}$. Therefore the correlations are not restricted only to power law fluids obeying the correlation $\tau = K\dot{\gamma}^n$ but can be applied to time-independent fluids following any rheological model. Dodge and Metzner found that their correlation fit their data best when the slope of the log-law $A = 1/\kappa$, where $\kappa$ is called Karman's universal constant, is expressed as $A = 2.5/n'^{0.75}$. Tennekes (1966) argued that the Dodge and Metzner correlation implied that the mechanism of turbulence in non-Newtonian fluids is different from that in Newtonian fluids. Dodge and Metzner also proposed an empirical extension to the Blasius (1913) power law correlation. The Dodge and Metzner correlations have been widely accepted from the moment they were published and are routinely quoted in books on non-Newtonian fluid technology e.g. (Chabra & Richardson, 1999; Skelland, 1967; Steffe, 1996) and remain highly recommended even in recent evaluations of correlations for friction factors in power law fluids e.g. Gao and Zhang (2007).

However, the predictions of velocity profiles proposed by Dodge & Metzner, based on the success of their friction factor correlation, did not agree with the subsequent measurements of Bogue and Metzner (1963) where the slope of the log-law is described by $A = 2.5/n'$. This discrepancy between results of researchers in the Metzner group has encouraged others to attempt to proposed improved correlations notably by Thomas (1960), Torrance (1963), Kemblowski and Kolodzieski(1973), Hancks and Ricks (1975), Szilas et al.(1981), Shenoy and Saini(1982), Kawase et al (1994), Wilson and Thomas(1985), Darby (1988), Desouky (Desouky, 2002; 1990), Hemeida (1993) and El-Emam et al. (2003)as shown in Table 1. Many of these correlations are empirical or semi-empirical. Indeed El-Amam et al tested 11 correlations against published data and noted that many correlations fitted the experimental data of their authors well enough but not that of others indicating a lack of generality. El-Amman et al. (op. cit.) proposed their own empirical correlation with fitted most of the literature data they used but the improved correlation gave no clue on the underlying mechanisms.

Wilson and Thomas (1985) proposed that the normalised Kolmogorov energy dissipating eddies (Kolmogorov, 1941) were larger in power law fluids than in Newtonian fluids by a factor of 2/(n+1) because of the integration of the shear rate into the local velocity $u = \int \dot{\gamma} dy = \int (\tau/K)^{1/n} dy$. They proposed that the scale of the Kolmogorov eddies normalised with the friction velocity $u_* = \sqrt{\tau_w/\rho}$ and the apparent viscosity at the wall $\mu_w = \tau_w^{1-1/n}/K^{1/n}$ is $\lambda^+ = \lambda u_* \rho/\mu_w = 12.1[2/(n+1)]$. However, their prediction of friction factors fell 5% to 15% below the well-accepted measurements of Dodge. A particular weakness of the Wilson and Thomas integration procedure is that it does not involve boundary conditions and therefore delivers the same result for different geometries like cylindrical pipes, parallel plates and annuli. These authors only tested their theory against pipe flow. Fifteen years earlier Trinh (1969) had also argued that the integration of the wall shear rate resulted in a different value for the intercept between the log-law and the curve representative of purely viscous flow, obeying the relation $u^+ = y^+$ with $u^+ = u/u^*$, was different in Newtonian and non-Newtonian flows. He argued that the result of the integration process was the same as that obtained in the widely used Mooney-Rabinowitsch procedure(Skelland, 1967) The shift factor for pipe flow of power law fluid is therefore $(3n'+1)/(4n')$. This correlation predicted the data of Dodge and Bogue with accuracy similar with the Metzner-Dodge correlation. Trinh also showed that the technique could be applied easily to other rheological models and that the shift factor also correlated very well the delay in the transition to turbulence when dimensionless flow rate was expressed as the Metzner-Reed Reynolds number. This work was never published because the author returned to Vietnam, a country then at war.

Table1. A number of friction factor correlations for pipe flow of power law fluids

| No. | Authors | Equation | Year |
|---|---|---|---|
| 1 | Dodge and Metzner | $1/\sqrt{f} = 4.0/(n')^{0.75} \log[\text{Re}_{MR} f^{1-n'/2}] - 0.4/(n')^{1.2}$ | 1959 |
| | | $f = a/\text{Re}_{MR}^b$ where $a=0.066\,5+0.011\,75n'$ and $b = 0.365 - 0.177n' + 0.062n'^2$ | |
| 2 | Shaver and Merilll | $f = 0.079/(n^5 \text{Re}_{MR}^\alpha), \alpha = 2.63/10.5^n$ | 1959 |
| 3 | Tomita | $1/\sqrt{f_{To}} = 4\log(\text{Re}_{To}\sqrt{f_{To}}) - 0.4$ | 1959 |
| | | $f_{To} = (4/3)[(1+2n)/(1+3n)]f$ | |
| | | $\text{Re}_{To} = (3/4)[(1+3n)/(1+2n)]\text{Re}_{MR}$ | |
| 4 | Thomas | $1/\sqrt{f} = 4/n' \log(\text{Re}_{MR} f^{1-n'/2}) - 0.4n'$ | 1960 |
| 5 | Clapp | $1/\sqrt{f} = (4.53/n')\log[\text{Re}_{MR} f^{1-n'/2}] + 2.69/n' + 0.68(5n' - 8/n')$ | 1961 |
| 6 | Trinh | $1/\sqrt{f} = (4.06/n')\log(\text{Re}_{MR} f^{1-n'/2}) + 2.16 - 2.78/n'$ | 1969 |
| | | $f = \alpha/\text{Re}_{MR}^\beta,$ | |
| | | $\beta = 1/(3n'+1)$ | |
| | | $\alpha = \left[11.8^{-6n'/(3n'+1)}((3n'+1)/4n')^{-7n'/(3n'+1)} 2^{(4+n')/(3n'+1)}\right]/\phi^{7n'/(3n'+1)}$ | |
| | | $\phi = V/U_{max}$ | |

| 7 | Kemblowski and Kolodziejski | $f = 0.0025 \exp(3.57n^2) \exp[572(1-n^{4.2})/n^{0.453} \text{Re}_{MR}] / \text{Re}_{MR}^{0.314n^2}$<br>$\text{when } \text{Re}_{MR} > 31600/n^{0.435}, \text{then } f = 0.079(1/\text{Re}_{MR}^{0.25})$ | 1973 |
|---|---|---|---|
| 8 | Hanks and Ricks | $f = 0.0682 n^{-1/2} / \text{Re}_{MR}^{1/(1.87+2.39n)}$ | 1973 |
| 9 | Shenoy and Saini | $1/\sqrt{f} = 3.57 \log\left[\text{Re}_{MR}^{1/n^{0.615}} / 6.5^{1/n^{1+0.75n}}\right]$ | 1986 |
| 10 | Desouky and El-Emam | $f = 0.125 n^{\sqrt{n}} (0.0112 + \text{Re}_{MR}^{0.3185})$ | 1990 |
| 11 | El-Eman et al | $f = \left[n'/(3.072 - 0.1433n') \text{Re}_{MR}^{n'/(0.282-4.211n')} - 0.00065\right]/4$ | 2003 |

There has been little interest to revisit the problem analytically in the last decade and recent research has focussed more on numerical computations e.g. (Malin, 1997; Rudman & Blackburn, 2006; Rudman, Blackburn, Graham, & Pullum, 2004)) or the use of neural networks (Sablani & Shayya, 2003)

This paper introduces a new visualisation of turbulence in time-independent non-Newtonian fluids and aims to answer a fundamental question: Are the mechanisms of turbulence in Newtonian and time-independent fluids different?

**Theory**

The velocity profile in turbulent flow is traditionally divided into two regions e.g. (Panton, 1990): a wall layer where the effects of viscosity are important, and an outer region where they are not. The portion of the velocity profile in the outer region adjacent to the wall region obeys a logarithmic relationship. The slope of the log-law sub-region was the same for all fluids, Newtonian and non-Newtonian (Bogue, 1961; Trinh, 1969; Trinh, 1994; Trinh, 2005).

**Structure of the wall layer**

The process in the wall layer of Newtonian turbulent flow was first observed by Kline, Reynolds, Schraub, & Runstadler (1967). Despite the dominating influence of viscous momentum, the flow field near the wall is not laminar in the steady-state sense, but highly active. Periodically, fast fluid rushes from the outer region towards the wall then follows a vortical sweep along the wall The travelling vortex induces underneath its path (Figure 1) which is observed as streaks of low-speed fluid. The streaks tend to lift, oscillate and eventually burst in violent ejections from the wall towards the outer region. The low speed streak phase is much more persistent than the ejection phase and dominates the contribution to the time-averaged profile (Walker, Abbott, Scharnhorst, R.K., & Weigand, 1989). This inrush-sweep-burst cycle is now regarded as central to the production of turbulence near a wall.

Einstein and Li (1956)were the first to propose that the intermittent wall layer should be modelled with as an unsteady state developing viscous layer in constrast to Prandtl's concept of a steady state viscous sublayer. They used the Stokes' solution for an impulsively started flat plate.with the governing equation

$$\frac{\partial u}{\partial t} = -\frac{1}{\rho}\frac{\partial \tau}{\partial y} \tag{1}$$

Many authors have subsequently used the Einstein-Li approach with further refinements to model the wall layer e.g. (Black, 1969; Hanratty, 1956, 1989; Meek & Baer, 1970). Reichardt (1971) has included the effect of the pressure gradient

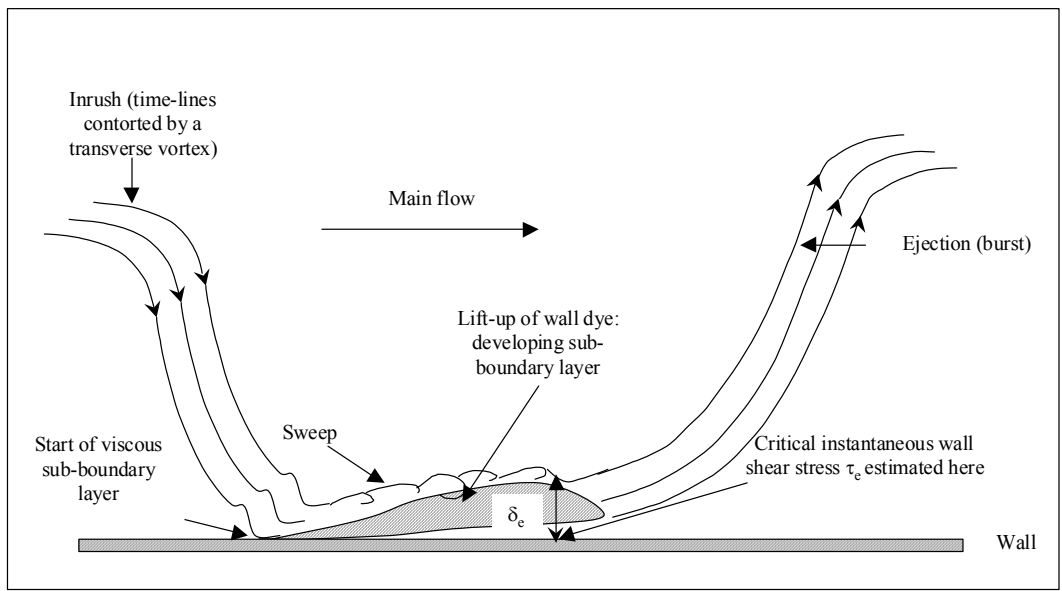

Figure 1. A cycle of the wall layer process. Drawn after the observations of Kline *et.al.* (1967)

Many researchers could not reconcile the concept of a laminar wall-layer, even intermittent, with the intense activity that Kline et al. (1967)first identified in the wall layer and confirmed by many others (Corino & Brodkey, 1969; Kim, Kline, & Reynolds, 1971; Offen & Kline, 1975). In particular the Stokes solution looked incompatible with the coherent structures that dominated the studies of turbulence in the last fifty years e.g.(Adrian, 2007; Cantwell, 1981; Carlier & Stanislas, 2005; Jeong & Hussain, 1995; Jeong, Hussain, Schoppa, & Kim, 1997; Robinson, 1991; Smith & Walker, 1995; Swearingen & Blackwelder, 1987). Indeed even a cursory search in The Web of Knowledge database returned thousands of papers devoted to coherent structures and entire books have been devoted to their understanding e.g. (Holmes, Lumley, & Berkooz, 1998). Particular attention has been paid to the horseshoe or hairpin vortices that have been seen by many as crucial to an understanding of wall turbulence e.g. (Arcalar & Smith, 1987b; Gad-el-Hak & Hussain, 1986; Schoppa & Hussain, 2000; Suponitsky, et al., 2005).and one researcher (McNaughton & Brunet, 2002)) stated the common "hope that understanding these 'coherent structures' will give insight into the mechanism of turbulence, and so useful information for explaining phenomena and formulating models." In the same breadth he acknowledged that "Unfortunately little of practical value has been achieved in the 50 years of research into turbulence structure because of the very complexity of turbulence, so that there is still no accepted explanation of what the observed structures are and how they are formed, evolve and interact."

**Statement of theory**

A better understanding is obtained by decomposing the instantaneous velocity in the wall layer Reynolds (1895) proposed that the instantaneous velocity $u_i$ at any point may be decomposed into a long-time average value $U_i$ and a fluctuating term $U'_i$.

$$u_i = U_i + U'_i \qquad (2)$$

The advance in measuring techniques of the last fifty years have shown conclusively that the instantaneous velocity traces of flow close to a wall show two types of fluctuations: fast and slow. Figure 1 shows a typical trace of streamwise velocity near the wall, redrawn after the measurements of (Antonia, Bisset, & Browne, 1990). If we draw a smooth line through this velocity trace so that there are no secondary peaks within the typical timescale of the flow $t_v$, we define a locus of smoothed velocity $\tilde{u}_i$ and fast fluctuations $u'_i$ of period $t_f$ relative this base line.

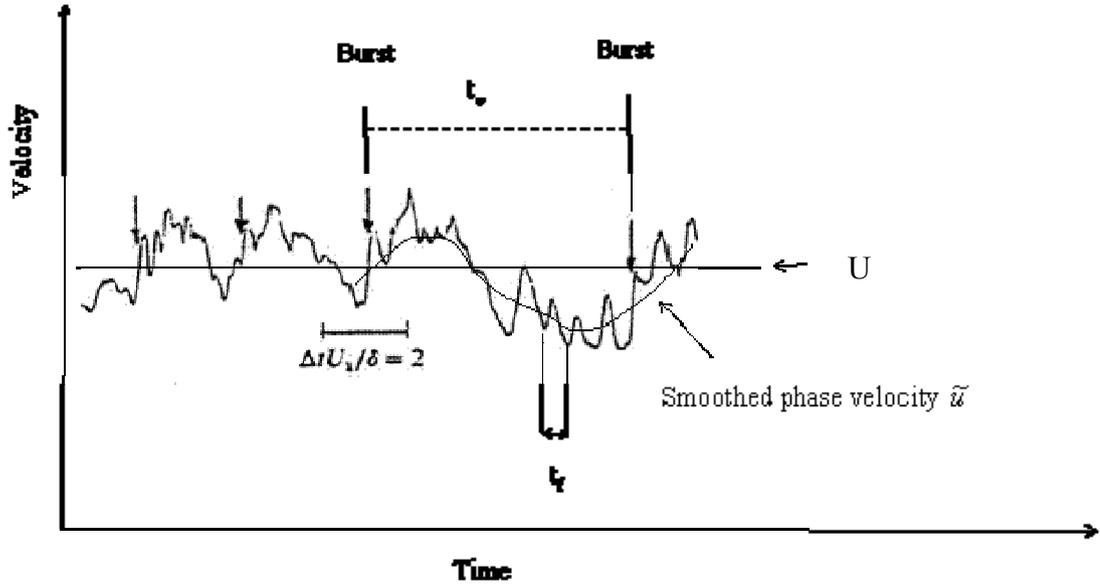

Figure 1 Trace of instantaneous streamwise velocity after measurements by Antonia et al. (1990).

The instantaneous velocity may be decomposed in an alternate manner as:

$$u_i = \tilde{u}_i + u'_i \tag{3}$$

Then we may write

$$\int_0^\infty u'_i \, dt = 0 \tag{4}$$

$$U'_i = \tilde{U}'_i + u'_i \tag{5}$$

where

$$\tilde{U}'_i = \tilde{u}_i - U_i \tag{6}$$

then

$$u_i = U_i + \tilde{U}'_i + u'_i \tag{7}$$

We may average the Navier-Stokes equations over the period $t_f$ of the fast fluctuations. Bird, Stewart, & Lightfoot (1960, p.158) give the results as

$$\frac{\partial(\rho \tilde{u}_i)}{\partial t} = -\frac{\partial p}{\partial x_i} + \mu \frac{\partial^2 \tilde{u}_i}{\partial x_j^2} - \frac{\partial \tilde{u}_i \tilde{u}_j}{\partial x_j} - \frac{\partial \overline{u'_i u'_j}}{\partial x_j} \tag{8}$$

Equation (8) defines a second set of Reynolds stresses $\overline{u'_i u'_j}$ which we will call "fast" Reynolds stresses to differentiate them from the standard Reynolds stresses $\overline{U'_i U'_j}$. In general $u'_i < U'_i$ and the fast Reynolds stresses are smaller in magnitude than the standard Reynolds stresses.

Within a period $t_\nu$, the smoothed velocity $\tilde{u}_i$ varies slowly with time but the fluctuations $u'_i$ may be assumed to be periodic with a timescale $t_f$. We may write the fast fluctuations in the form

$$u'_i = u_{0,i}\left(e^{i\omega t} + e^{-i\omega t}\right) \qquad (9)$$

The fast Reynolds stresses $u'_i u'_j$ become

$$u'_i u'_j = u_{0,i} u_{0,j}\left(e^{2i\omega t} + e^{-2i\omega t}\right) + 2 u_{0,i} u_{0,j} \qquad (10)$$

Equation (10) shows that the fluctuating periodic motion $u'_i$ generates two components of the "fast" Reynolds stresses: one is oscillating and cancels out upon long-time-averaging, the other, $u_{0,i} u_{0,j}$ is persistent in the sense that it does not depend on the period $t_f$. The term $u_{0,i} u_{0,j}$ indicates the startling possibility that a purely oscillating motion can generate a steady motion which is not aligned in the direction of the oscillations. The qualification steady must be understood as independent of the frequency ω of the fast fluctuations. If the flow is averaged over a longer time than the period $t_\nu$ of the bursting process, the term $u_{0,i} u_{0,j}$ must be understood as transient but non-oscillating. This term indicates the presence of transient shear layers embedded in turbulent flow fields and not aligned in the stream wise direction similar to those associated with the streaming flow in oscillating laminar boundary layers (Schneck & Walburn, 1976; Tetlionis, 1981). Schoppa and Hussain(2002)also observed that sinusoidal velocity fluctuations in turbulent flows led to the production of intense shear layers associated with the streaming flow, that they call transient stress growth TSG.

Thus to describe the process of the wall layer we need to superimpose two solutions: Stokes' solution for a flat plate suddenly set in motion that we will call here solution1 and Stokes'solution for an oscillating flat plate that we call solution2.

We first look at the effect of the fast fluctuations by analysing an oscillating flow with a zero-mean velocity . In this case one may thus investigate the effect of the amplitude and frequency of the fluctuations separately because the basic velocity fluctuations imposed by external means do not grow with time.. The following treatment of the problem is taken from the excellent book of (Tetlionis, 1981).

We define a stream function $\psi$ such that

$$u = \frac{\partial \psi}{\partial y} \qquad v = \frac{\partial \psi}{\partial x} \tag{11}$$

Where $u$ and $v$ are the components o local instantaneous velocity in the x, y directions. The basic variables are made non-dimensional

$$x^* = \frac{x}{L} \qquad y^* = \frac{y}{\sqrt{2\nu/\omega}} \qquad t^* = t\omega \tag{12}$$

$$U_e^*(x,t) = \frac{U_e}{U_\infty}(x,t) \qquad \psi^* = \psi \left( U_\infty \sqrt{\frac{2\nu}{\omega}} \right)^{-1} \tag{13}$$

where $U_\infty$ is the approach velocity for $x \to \infty$, $U_e$ is the local mainstream velocity and L is a characteristic dimension of the body. The system of coordinates x, y is attached to the body. The Navier-Stokes equation may be transformed as:

$$\frac{\partial^2 \psi^*}{\partial y^* \partial t^*} - \frac{1}{2} \frac{\partial^3 \psi^*}{\partial y^{*3}} - \frac{\partial U_e^*}{\partial t^*} = \frac{U_e}{L\omega} \left( -\frac{\partial \psi^*}{\partial y^*} \frac{\partial^2 \psi^*}{\partial y^* \partial x^*} + \frac{\partial \psi^*}{\partial x^*} \frac{\partial^2 \psi^*}{\partial y^{*2}} + U_e^* \frac{\partial U_e^*}{\partial x^*} \right) \tag{14}$$

with boundary conditions

$$\psi^* = \frac{\partial \psi^*}{\partial y^*} = 0 \qquad y^* = 0 \tag{15}$$

For large frequencies, the RHS of equation (14) can be neglected since

$$\varepsilon = \frac{U_e}{L\omega} \ll 1 \tag{16}$$

In this case, Tetlionis reports the solution of equation (14) as:

$$\psi^* = \left[ \frac{U_0^*(x^*)}{2}(1-i)[1 - e^{(1+i)y^*}] + \frac{U_0^* y^*}{2} \right] e^{it^*} + C \tag{17}$$

Tetlionis (op. cit. p. 157) points out that equation (17) may be regarded as a generalisation of Stokes' solution (1851) for an oscillating flat plate.

Equation (17) is accurate only to an error of order $\varepsilon$. Tetlionis reports a more accurate solution for the case when $\varepsilon$ cannot be neglected (i.e. for lower frequencies):

$$\psi^* = \frac{U_0^*(x^*)}{2}[\psi_0^*(y^*)e^{it^*}\overline{\psi_0^*(y^*)e^{-it^*}}] + \varepsilon[\psi_1^*(x^*,y^*)e^{2it^*} + \overline{\psi_0^*}e^{-2it^*}] + O(\varepsilon^2) \qquad (18)$$

where $\psi_0$ and $\psi_1$ are the components of the stream function of order $\varepsilon^0$ and $\varepsilon$. Substituting this more accurate solution into equation (14), we find that the multiplication of coefficients of $e^{it^*}$ and $e^{-it^*}$ forms terms that are independent of the oscillating frequency, $\omega$, imposed on the flow field and were not anticipated in equation (18). Thus the full solution of equation (14) is normally written (Stuart, 1966; Tetlionis, 1981) as

$$\psi^* = \frac{U_0^*(x^*)}{2}[\psi_0^*(y^*)e^{it^*} + \overline{\psi_0^*(y^*)}e^{-it^*}]$$
$$+ \varepsilon[\psi_{st}^* + [\psi_1^*(x^*,y^*)e^{2it^*} + \overline{\psi_1^*(x^*,y^*)}e^{-2it^*}] + O(\varepsilon^2) \qquad (19)$$

where the overbar denotes the complex conjugate and $\psi_{st}^*$ results from cancelling of $e^{it^*}$ and $e^{-it^*}$ terms.

The quantity $\psi_{st}^*$ shows that the interaction of convected inertial effects of forced oscillations with viscous effects near a wall results in a non-oscillating motion that is referred to in the literature as "Streaming". The problem has been known for over a century (Faraday, 1831; Dvorak, 1874; Rayleigh, 1880, 1884; Carriere, 1929; Andrade, 1931; Schlichting, 1932) and studied theoretically (Riley, 1967; Schlichting, 1960; Stuart, 1966; Tetlionis, 1981).

When the smoothed phase velocity cannot be neglected, the problem becomes much more complex. Early analytical investigations (Riley, 1975; Stuart, 1966)) focused mainly on laminar oscillating boundary layers but could not capture all the elements of the flow field because of many simplifications and omissions were required to overcome the considerable mathematical difficulties. They did give insight into the role of different terms in the NS equations. Numerical solutions (Schneck & Walburn, 1976) could capture the patterns of viscous and streaming flow but gave less insight into the role of the different terms in the NS equations. Yet more information was provided in the rational numerical simulations RNS (nomenclacutre of Zhang (1991)) that study the interactions between the different structures embedded in a turbulent flow field. In particular, the effect of vortices as they move above the wall has been studied in a number of "kernels" (nomenclature of Smith et al (1991)). These kernel studies show that a vortex moving above a wall will induce a laminar sub-boundary layer underneath its path by viscous diffusion of momentum, even if the vortex is introduced into a fluid which was originally at rest (Smith, Walker, Haidari, & Sobrun, 1991). The vortex impresses a periodic disturbance onto the laminar sub-boundary layer underneath. which oscillates then erupts in a violent ejection that Peridier et al (1991) call viscous-inviscid interaction The problem is thus very similar to the streaming flow discussed by Tetlionis. In these kernel studies the configuration of the vortex must be specified a priori. In the work of Walker (1978) it is a rectilinear vortex, in Chu and Falco (1988) ring vortices, in Liu et al. (1991) hairpin vortices, in Swearingen and Blackwelder (1987), streamwise Goertler vortices. But recently in their numerical simulation Suponitsky, Cohen, & Bar-Yoseph (2005) have shown that vortical disturbances evolve into a hairpin vortex independently of their original geometry over a wide range of orientations. The eruptions are linked with the growth of the vortex as it moves down the wall, seen clearly in hydrogen bubble visualisations e.g. (Offen & Kline, 1974)and the growth of the fast velocity fluctuations impressed on the viscous sub-boundary layer. It occurs at a critical value of the term $\varepsilon$ in equation (16), which is a combination of the Reynolds number of the sub-boundary layer and the Strouhal number of the fluctuations.Thus we can use the Stokes solution1 which is a particular form of the solution of order $\varepsilon^0$ as Einstein and Li proposed to model the sweep phase of the wall layer but it must be expressed in terms of the smoothed velocity i

$$\frac{\partial \tilde{u}}{\partial t} = -\frac{1}{\rho}\frac{\partial \tau}{\partial y} \qquad (20)$$

A more detailed discussion of these issues in presented in (Trinh, 2009)

It is now proposed that the development of instabilities leading to ejections, in other words the bursting process, is the same for Newtonian and non-Newtonian fluids. More specifically, it is postulated that the ejections always occur when the ratio of kinetic to viscous energy in the transient sub-boundary layer reaches a critical value, which is not dependent on the nature of the fluid. This critical ratio can be estimated by a kind of local instantaneous friction factor. It is calculated from the critical instantaneous wall shear stress and the local approach velocity to the sub-boundary layer at the end of the low-speed streak phase, just prior to ejection (as shown in Figure 1).

Since direct measurements of this critical instantaneous shear stress are difficult, it is estimated from time-averaged measurements through a solution of equation (20). The analysis is made here for fluids that obey the Ostwald de Waele power law rheological model

$$\tau = K \dot{\gamma}^n \tag{21}$$

where K is called the consistency coefficient,

    n           the flow behaviour index, and

    $\dot{\gamma}$           the shear rate

Application to other fluids models is presented in subsequent publications.

Substituting equation (21) into (20) gives

$$\rho \frac{\partial \tilde{u}}{\partial t} = n K \left( \frac{\partial \tilde{u}}{\partial y} \right)^{n-1} \frac{\partial^2 \tilde{u}}{\partial y^2} \tag{22}$$

The relative distance $\eta(t, y)$ into the viscous sub-boundary layer near the wall $\delta_i(t)$ is defined as

$$\eta(t, y) = \frac{y}{\delta_i(t)} \tag{23}$$

and the velocity $\phi$ relative to the velocity $\tilde{U}_e$ at the edge of the wall layer is

$$\phi = \frac{\tilde{u}}{\tilde{U}_e} \tag{24}$$

Substituting these new variables into equation (22) and integrating with respect to η gives

$$\delta_i^n \frac{d \delta_i}{dt} = \frac{K}{\rho} \tilde{U}_e^{n-1} \frac{N}{M} \tag{25}$$

in which

$$N = \int_0^1 n(\phi')^{n-1}(\phi'') d\eta(t,y) = [(\phi')^n]_0^1 \tag{26a}$$

and

$$M = \int_0^1 (\phi')\eta(t,y)d\eta(t,y) = 1 - \int_0^1 \phi \, d\eta(t,y) \qquad (26b)$$

where the primes denote derivatives with respect to η.

Equation (25) can now be integrated separately with respect to the variables $\delta_i$ and t over a characteristic time $t_v$:

$$\int_0^{\delta_e} \delta_i^n \, d\delta_i = \int_0^{t_v} \frac{K}{\rho} \tilde{U}_e^{n-1} \frac{N}{M} dt \qquad (27)$$

The instantaneous wall-layer thickness at time $t_v$, which coincides with the onset of ejection, is

$$\delta_e = \delta_i(t_v) = \left[ \frac{K}{\rho}(n+1)\tilde{U}_e^{n-1} \frac{N}{M} t_v \right]^{1/(n+1)} \qquad (28)$$

The instantaneous wall shear stress is:

$$\tau_{w,i} = K \left[ -\left( \frac{\partial \tilde{u}}{\partial y} \right)_{w,i} \right]^n \qquad (29a)$$

$$\tau_{w,i} = K \left[ \tilde{U}_e (\phi')_w \frac{\partial \eta}{\partial y} \right]^n \qquad (29b)$$

$$\tau_{w,i} = KN \frac{\tilde{U}_e^n}{[\delta_i(t)]^n} \qquad (29c)$$

The time-averaged wall shear stress is given by

$$\tau_w = \frac{1}{t_v} \int_0^{t_v} \tau_{w,i} \, dt \qquad (30a)$$

$$\tau_w = K \, N \, (n+1) \frac{\tilde{U}_e^n}{\delta_e^n} \qquad (30b)$$

$$\tau_w = (n+1)\tau_e \qquad (30c)$$

where $\tau_e = \tau_{w,T}$ is the instantaneous wall-shear stress at time $t_v$ which coincides with the end of the low-speed-streak phase and the onset of ejection, as shown in Figure 1. Henceforth, this wall shear stress will be called the *critical local instantaneous wall-shear stress at the point of ejection* or simply the *critical shear stress*.

The transient unsteady state sub-boundary layer has been called a Stokes layer and can often be found embedded in other flows (Tetlionis, 1981). The thickness of the Stokes layer $\delta_e$ at the end of the period $t_v$ is related to the critical approach velocity $\tilde{U}_e$ by putting $t = t_v$ in equation (29c) and rearranging:

$$\delta_e = \left[\frac{KN}{\tau_e}\right]^{1/n} \tilde{U}_e \qquad (31)$$

The time-averaged shear velocity is usually given the symbol $u_*$ and defined as:

$$u_* = \sqrt{\frac{\tau_w}{\rho}} \qquad (32)$$

We define a new normalising parameter, the *critical* shear velocity $u_{e*}$

$$u_{e*} = \sqrt{\tau_e/\rho} = \sqrt{\frac{\tau_w}{\rho(n+1)}} = \frac{u_*}{\sqrt{n+1}} \qquad (33)$$

The thickness of the Stokes layer may be normalised with the critical wall shear stress as:

$$\delta_e^+ = \frac{\delta_e u_{e*}}{\nu_e} = \frac{\delta_e u_{e*}^{2/n-1} \rho^{1/n}}{K^{1/n}} \qquad (34)$$

The normalised critical approach velocity at the point of bursting is

$$\tilde{U}_e^+ = \frac{\tilde{U}_e}{u_{e*}} \qquad (35)$$

Combining equations (31), (33), (34) and (35) gives

$$\delta_e^+ = N^{1/n} \tilde{U}_e^+ \qquad (36)$$

The coefficients M and N can be determined once the relative velocity $\phi$ as a function of $\eta$ is known. Following Polhausen (1921) and Bird, Stewart, & Lightfoot (1960), we assume that the velocity profile can be described approximately by a third-order polynomial:

$$\phi = 1.5\eta - 0.5\eta^3 \qquad (37)$$

Back-substitution into equations (26a) and (26b) respectively yields the unknown

coefficients:

$$N = (3/2)^n \tag{38}$$

and

$$M = 3/8 \tag{39}$$

Substitution of equation (38) into (36) gives

$$\delta_e^+ = 1.5 \tilde{U}_e^+ \tag{40}$$

Equation (40) shows clearly that the relation between the wall layer thickness and the critical approach velocity of the Stokes layer at the point of bursting is independent of the fluid rheology, specifically the flow behaviour index, when normalised with the critical instantaneous shear velocity.

We should note that for Newtonian fluids, n=1, equation (38) gives

$$N_{n=1} = 3/2 = (N)^{1/n} \tag{41}$$

Therefore equation (40) may be written as:

$$\delta_e^+ = N_{n=1} \tilde{U}_e^+ \tag{42}$$

Thus the previous conclusion is not dependent on the form of the velocity profile assumed. For example, the same derivation leading to equation (42) may be performed with a fourth-order polynomial also proposed by Polhausen (1921). Of course the numerical value of the coefficient N changes with the velocity profile assumed but equation (40) does not. In the exact Stokes solution (Bird, 1959)

$$N_{n=1} = 2.08 \tag{43}$$

Because instantaneous shear and velocity profiles are very difficult to measure, it is convenient to re-express these instability criteria in terms of time-averaged shear stresses through the use of equation (30c). Since the edge of the wall layer is defined by the maximum penetration of viscous momentum from the wall then the time averaged wall layer thickness is equal to the thickness of the transient viscous sub-boundary layer at the point of ejection $\delta_v = \delta_e$ and $U_v = \tilde{U}_e$. Then the velocity at the edge of the wall layer, normalised with the time-averaged shear velocity, becomes

$$U_v^+ = \frac{\tilde{U}_e}{\sqrt{\tau_w/\rho}} = \frac{\tilde{U}_e^+}{\sqrt{n+1}} \tag{44}$$

Similarly

$$\delta_v^+ = \frac{\delta_e u_*^{2/n-1} \rho^{1/n}}{K^{1/n}} = \delta_e^+ (n+1)^{2-n/2n} \tag{45}$$

Combining equations (42), (44) and (4527) gives

$$\delta_v^+ = 2.08(n+1)^{1/n} U_v^+ \tag{46}$$

Equations (42) and (46) indicate that the apparent thickening of the wall layer, seen in traditional velocity plots normalised with the time-averaged shear velocity $u_*$, such as those of (Bogue (1961), are not real. This apparent thickening is the consequence of an integration process, which relates the critical instantaneous wall shear stress, the normalising parameter with physical significance, to the time-averaged wall shear stress, which is traditionally more easily measured.

**Friction factors and Reynolds numbers**

The critical apparent (non-Newtonian) kinematic viscosity at the wall $v_e$ is defined as

$$v_e = \frac{\tau_e}{\rho(-\partial \tilde{u}/\partial y)_e} = K^{1/n} \tau_e^{(n-1)/n} \tag{47}$$

The critical instantaneous Reynolds number becomes

$$Re_e = \frac{DV}{v_e} = \frac{DV}{K^{1/n} \tau_e^{(n-1)/n}} \tag{48}$$

and the critical instantaneous friction factor is

$$f_e = \frac{2\tau_e}{\rho V^2} \tag{49}$$

These definitions can be compared with the more conventional definitions of the time-averaged friction factor

$$f = \frac{2\tau_w}{\rho V^2} \tag{50}$$

and the generalised Metzner-Reed Reynolds number (Metzner & Reed, 1955)

$$Re_g = \frac{D^n V^{2-n} \rho}{K 8^{n-1} \left(\frac{3n+1}{4n}\right)^n} \tag{51}$$

The relation between the critical instantaneous Reynolds number and the generalised Metzner-Reed Reynolds number can be derived from equations (30c), (48), (49), (50) and (51):

$$\text{Re}_e = \left(\text{Re}_g f^{1-n}\right)^{\frac{1}{n}} 2^{\frac{5(n-1)}{n}} \left(\frac{(3n+1)}{4n}\right)\left(\frac{n+1}{2}\right)^{\frac{n-1}{n}} \tag{52}$$

**Verification of theory**

The friction factor in viscous non-Newtonian pipe flow has been measured by Dodge (1959), Bogue (1961) and (Yoo, 1974). Figure 3 shows a plot of time-averaged friction factor against the Metzner-Dodge Reynolds number.

The data for different values of the flow behaviour index, n, falls on different lines. Figure 4 shows a plot of the critical instantaneous friction factor against the critical instantaneous Reynolds number, defined in terms of the critical wall shear stress. All the data falls on a unique plot.

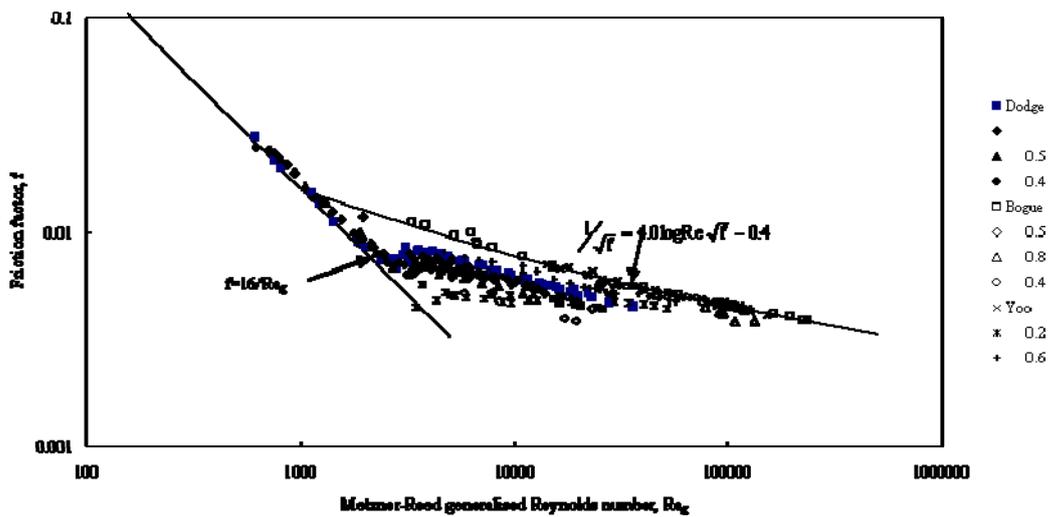

Figure 3. Plot of time-averaged friction factor against generalized Metzner-Reed Reynolds number.

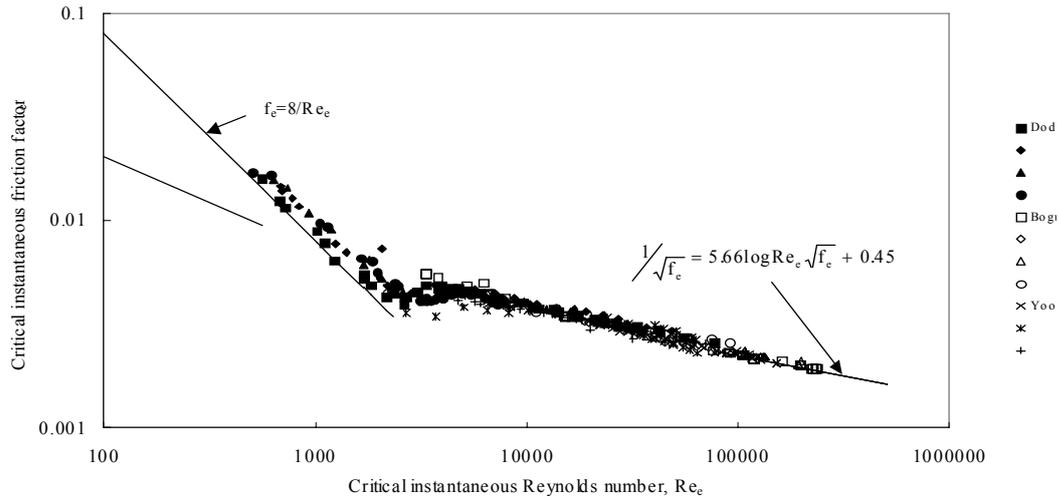

Figure 4. Similarity plot of instantaneous critical friction factor against Reynolds number.

For Newtonian fluids, the critical friction velocity is related to the time averaged friction velocity by putting $n = 1$ in equation (33) and:

$$u_{e*} = u_* \sqrt{2} \tag{53}$$

Similarly equation (30c) gives

$$\tau_w = 2\tau_e \tag{54}$$

$$f = 2f_e \tag{55}$$

Prandtl's logarithmic law (Prandtl, 1935; Nikuradse, 1932)

$$\frac{1}{\sqrt{f}} = 4.0 \log(Re \sqrt{f}) - 0.4 \tag{56}$$

may be rewritten in terms of the critical wall shear stress

$$\frac{1}{\sqrt{f_e}} = 5.66 \log(Re_e \sqrt{f_e}) + 0.45 \tag{57}$$

Equation (57) fits the data of Dodge, Bogue and Yoo quite closely, as shown in Figure 4. The agreement is not perfect because the wall layer analysis has been made for flow past a flat surface. The application of this analysis to circular pipe flow implies that the curvature effects can be neglected, which is true only at high Reynolds numbers when the wall layer is very thin compared to the pipe radius.

Similarly, the Blasius power law (Blasius, 1913)

$$f = \frac{0.079}{Re^{1/4}} \tag{58}$$

may be rewritten as

$$f_e = \frac{0.04}{Re_e^{1/4}} \tag{59}$$

Equation (59) also fits the data presented in Figure 4.

**Discussion and Conclusion**

The apparent viscosity of non-Newtonian fluids changes with the applied shear stress or shear rate. Therefore, it is important for engineering correlations to identify exactly the shear stress (or shear rate) at which the viscosity should be evaluted. It can then be used to normalise the velocity $u$, the distance y and to calculate the Reynolds number. Experimental data shows that both the velocity profile (e.g. (D. C. Bogue & Metzner, 1963)) and the friction factor-Reynolds number curves (Figure3) of power law fluids are shifted when the velocity and distance are normalised with the viscosity calculated at the time averaged value of the wall shear stress. The data gathered by Bogue (D.C. Bogue, 1961) and others e.g.(Trinh, 2005) show that the wall thickness $\delta_v^+$, normalised with the time averaged wall shear stress, becomes thicker as the behaviour index n decreases. In an effort to collapse the non-Newtonian data onto their Newtonian counterpart many authors have resorted to different definitions of effective viscosity (e.g. Acrivos et al., Metzner and Reed, Edwards and Smith op. cit.). The present paper argues that since turbulence is a time-dependent phenomenon, the viscosity should be estimated at the value of the local instantaneous wall shear stress, not the time averaged value. When this is done, there is no need to define an effective viscosity. Even in laminar, Trinh and Keey(Trinh & Keey, 1992) argued that the diffusion of viscous momentum from the wall into the main flow is a time-dependent phenomenon: elements of fluid are convected along the main direction of flow and viscous momentum is diffused between adjacent fluid particles. Therefore the front of diffusion of viscous momentum moves across the pipe or laminar boundary layer only once but the process repeats itself regularly so that the time averaged profile looks steady.

The collapse of non-Newtonian laminar critical instantaneous friction factors of power law fluids onto the Poiseuille solution is not perfect because the Stokes solution was derived for an impulsively started flat plate and does not account for the curvature of the wall in pipe flow. This neglect is not important in turbulent flow when the wall layer is thin but becomes important for laminar flow that covers the entire pipe radius. An unsteady solution that accounts for curvature has been derived by Szymansky(Szymanski, 1932) and can be applied to models of turbulent flow ((Trinh, 1992; Trinh, 2009).) but the solution involves a Bessel series which is more cumbersome to handle mathematically.

The collapse of all turbulent friction factors expressed in terms of the critical instantaneous wall shear stress into a single curve suggests strongly that the mechanism of turbulence is the same for both Newtonian and power law fluids. In fact the same exercise can be repeated for other fluid models, for example the Bingham plastic and Herschel-Bulkley models, and the observation can be generally applied to all time-independent fluids. The shift observed when the time averaged shear stress is used as a normalising parameter arises from an integration constant and does not relate to the behaviour of polymer molecules in turbulent flow as sometimes postulated (e.g. (Wilson & Thomas, 1985)) or a pseudo-slip effect (Kozicki & Tiu,(1967) similar to the well documented observations in rheometric measurements of particulate suspensions (Steffe, 1996). An exception must be made for viscoelastic fluids because the elasticity of the polymers do dampen the periodic velocity fluctuations impressed on the wall layer by the travelling vortex ((Stone, Roy, Larson, Waleffe, & Graham, 2004; Trinh, 1969; Trinh, 1992; Trinh, 2009)) and reduce the value of the transient turbulent stresses responsible for the bursts.

In another report that summarises the writer's body of work on turbulence over a period of 40 years it is shown that the velocity profiles in the inner region (wall layer + log law region) of turbulent flows of Newtonian and non-Newtonian fluidsin many geometrical situations all collapse into a unique master curve when normalised with $U_v^+$ and $\delta_v^+$ (Trinh, 2009),p.84). The same similarity analysis also collapses the skin drag friction factors of Newtonian and non-Newtonian fluids unto a uniques master curve (Trinh, 2009), p.160). This evidence further supports the main argument presented in this paper: that the mechanism of turbulence is the same in Newtonian and time-independent fluids and that graphical differences arise from the different ways that the normalisation process is handled mathemathematically.

The present analysis has also demonstrated the existence of a similarity plot between the critical friction factor and the critical non-Newtonian Reynolds number for turbulent pipe flow of purely viscous power law fluids. Instabilities leading to the ejection of low-speed fluid from the wall layer are local phenomena and should be studied in terms of local instantaneous parameters. The critical point noted in this theory is that turbulence cannot be adequately explained by measurements and analysis of time-averaged shear stresses and velocities only. A great confusion has resulted from starting the theoretical analysis with the time-averaged Navier-Stokes equations, the Reynolds equations, whereas the present theory starts with the unsteady state Navier-Stokes equations and then time-averages the solution.

**Nomenclature**

| | |
|---|---|
| D | Pipe diameter |
| f | Time-averaged friction factor defined by equation (50) |
| $f_e$ | Critical (local instantaneous) friction factor defined by equation (49) |
| K | Consistency coefficient in power law model |
| M, N | Coefficients defined by equations (26a) and (26b) |
| n | Flow behaviour index in power law model |
| $Re_e$ | Critical instantaneous Reynolds number, $DV/\nu_e$ |
| $Re_g$ | Metzner-Reed generalised Reynolds number, equation (51) |
| t | Time |
| $\tilde{u}$ | Local instantaneous velocity smoothed with respect to fluctuations imposed by a travelling vortex |
| $u_*$ | Time-averaged friction velocity |
| $u_{e*}$ | Critical (instantaneous) friction velocity |
| $\tilde{U}_e$ | Approach velocity to transient viscous sub-boundary layer (velocity at wall layer edge at the moment of ejection) |
| $U_v^+$ | Approach velocity normalised with time-averaged friction velocity, $U_e/u_*$ |
| $\tilde{U}_e^+$ | Approach velocity normalised with the critical shear velocity, $\tilde{U}_e/u_{e*}$ |
| V | Mixing cup average discharge velocity |
| y | Normal distance from the wall |
| $\delta_i(t)$ | Instantaneous thickness of (Stokes) transient sub-boundary layer |
| $\delta_e^+$ | Wall layer normalised with the critical shear velocity |
| $\delta_v^+$ | Wall layer thickness normalised with the time-averaged shear velocity |
| $\dot{\gamma}$ | Shear rate |
| $\eta$ | Dimensionless distance defined by equation (23) |
| $\nu_e$ | Apparent critical kinematic viscosity defined by equation (47) |
| $\rho$ | Density |
| $\tau$ | Shear stress |
| $\phi$ | Dimensionless velocity defined by equation (24) |

Subscripts

| | |
|---|---|
| $\nu$ | Viscous, or at the edge of the wall layer |
| e | At the onset of ejection or at the end of the low-speed streak phase, see also T |
| i | instantaneous |
| $t_\nu$ | At the end of the low-speed-streak phase |
| w | Wall |

Superscript

| | |
|---|---|
| + | Normalised with wall shear stress |